 \documentclass[final,5p,times,twocolumn,authoryear]{elsarticle}

\usepackage{graphicx}
\usepackage{dcolumn}
\usepackage{bm}
\usepackage{amsmath}
\usepackage{amssymb}
\usepackage{lipsum}

\journal{Physics Letters B}

\bibpunct{[}{]}{,}{n}{}{,}  
\usepackage{orcidlink} 
\usepackage{booktabs}

\begin{document}

\begin{frontmatter}


\title{Accurate  spontaneous fission half-lives from a microscopic large-scale nuclear structure model}


\author[first,second]{A. Sánchez-Fernández}
\author[first,second,third]{S. Bara}
\author[first,second]{W. Ryssens}
\author[first,second]{S. Goriely}

\affiliation[first]{
  organization={Institut d'Astronomie et d'Astrophysique, Université Libre de Bruxelles},
  city={Brussels},
  country={Belgium}
}

\affiliation[second]{
  organization={Brussels Laboratory of the Universe - BLU-ULB},
  city={Brussels},
  country={Belgium}
}

\affiliation[third]{
  organization={KU Leuven, Instituut voor Kern- en Stralingsfysica},
  city={Leuven},
  postcode={3001}, 
  country={Belgium}
}

\begin{abstract}

We demonstrate that energy density functional (EDF) models for atomic nuclei can achieve a level of accuracy in describing fission properties comparable to that of more phenomenological approaches, while maintaining predictive power for nuclear ground-state observables. Using multiple BSkG parameterizations, we conducted a comprehensive benchmark of spontaneous fission (SF) half-lives and compare our predictions with all available experimental data - 136 values - from both ground states and fission isomers. Leveraging the quality of the BSkG functionals and explicitly accounting for more than two collective degrees of freedom in our calculations, we attained one of the closest agreements with experiment reported by an EDF model so far---within less than four orders of magnitude. This level of accuracy, together with the computational efficiency of our method, opens the way to systematic studies of the thousands of neutron-rich nuclei crucial for modeling r-process nucleosynthesis and the inclusion of SF half-lives into the fitting protocol of future EDF models.

\end{abstract}


\end{frontmatter}

 
Fission processes---whether spontaneous, $\beta$-delayed, neutron-induced or photon-induced---are arguably the most challenging type of nuclear reactions to model accurately: describing the split of a heavy nucleus into multiple fragments means facing simultaneously the complexities of the quantum many-body problem, the nucleon-nucleon interaction and out-of-equilibrium processes~\cite{Schunck2016,Bender20,Schunck2022}. Despite the difficulty, the need for hard-to-measure data in several fields of research continues to motivate improvements in fission theory: examples include societal challenges regarding energy and non-proliferation~\cite{Nupecc22,aliberti2006,sartori2013} and fundamental questions such as the stability of superheavy elements~\cite{nazarewicz2018,giuliani2019} and the reactor antineutrino anomaly~\cite{sonzogni2015,sonzogni2017,schmidt2021}. The study of heavy element nucleosynthesis by the rapid neutron-capture process (or r-process) is perhaps the most demanding application: not only do r-process simulations require estimates for the fission properties of thousands of extremely neutron-rich isotopes~\cite{arnould2020}, the predicted elemental abundances depend strongly on the efficiency of fission~\cite{martinez2007,goriely2015}.  The observational study of r-process nucleosynthesis is concerned too: fission processes - together with $\beta$-decay - essentially power kilonovae~\cite{metzger2010,giuliani2020}.

From this perspective, the key challenge in fission modelling is predictive power: how reliable can a model extrapolate to energies or neutron numbers that the (scarce) available data does not cover? A completely rigorous uncertainty quantification is likely only possible in an \textit{ab initio} context~\cite{Hergert2020}, but even the description of the ground state (GS) of the heavy nuclei of interest in fission studies is out of reach for these techniques. We are left with imperfect means to judge reliability: one should prefer models that capture as much as microscopic aspects of nuclei as is feasible, while describing the widest possible range of data accurately and \textit{simultaneously}.

Models based on an energy density functional (EDF) are the most promising avenue in this quest: state-of-the-art EDF-based models describe nuclear ground state properties across the entire nuclear chart with an accuracy that essentially matches that of more phenomenological microscopic-macroscopic (``mic-mac") models~\cite{Pomorski03,Goriely2013a,Scamps2021}. EDF-based modelling has also yielded important qualitative insights in fission processes~\cite{Delaroche06,marevic2021,Bulgac2022a}, but quantitative agreement with available fission data across large numbers of nuclei has so far been the domain of mic-mac models~\cite{randrup1976,moller2009,bao2013,moller2016,Blanco23}. Even for the latter, combining accuracy for GS properties and fission is not trivial: Ref.~\cite{moller2016} for example recommends different models when predicting masses or fission barriers.

In this Letter, we demonstrate that EDF-based modelling can \emph{simultaneously} describe GS properties and fission properties with
accuracy. More precisely, we demonstrate that calculations with the BSkG3 model \cite{Grams23}
reproduce all known spontaneous fission (SF) half-lives---107 values for GS and 29 for isomeric decay---within less than four orders
of magnitude, the accuracy expected of mic-mac models~\cite{moller1994,moller2009}. Our predictions are based solely on large-scale EDF calculations that account for three collective variables and do not involve additional parameters. This performance for SF half-lives adds
to the established quality of BSkG3 with respect to GS properties, reaching global root-mean-square (rms) deviations of 0.631 MeV and 0.024 fm for all known masses and charge radii, respectively~\cite{Grams23}. Besides thousands of masses, hundreds of charge radii and the properties of infinite nuclear matter, the parameter adjustment included all available empirical barriers of the RIPL-3 compilation and fission isomer excitation energies for nuclei with $Z\geq 90$~\cite{samyn2004,capote2009,Grams23}.  Although demanding, this procedure resulted in a model that combines excellent GS properties with an unsurpassed reproduction of the RIPL-3 barriers and available isomer excitation energies~\cite{capote2009}: BSkG3 reproduces the 45 primary and secondary barrier heights with rms errors of 0.33 MeV and 0.51 MeV, respectively, and 28 isomer excitation energies with an rms of 0.36 MeV~\cite{Grams23}. To the best of our knowledge, no other global approach - EDF-based or not - achieves such accuracy for both barriers and isomer excitation energies simultaneously~\cite{Ryssens23}. This performance reflects the quality of our adjustment strategy: BSkG2 \cite{ryssens2022} and BSkG4 \cite{,Grams25} offer comparable performance w.r.t.~barriers, isomeric excitation energies and SF half-lives, even if they differ from BSkG3 in important ways.


Since time-dependent simulations are prohibitively expensive when targeting many nuclei~\cite{jin2021},
we rely on the adiabatic approximation to reduce the complicated many-body dynamics of a fissioning nucleus
to a tunneling problem in a few collective coordinates~\cite{Brack1972,Baran1978,Schunck16,Bender20}. The semiclassical tunneling probability associated with a parameterized trajectory $L(s)$ in this collective space is given by

\begin{equation}
    P(L)=\left[1+e^{2S(L)}\right]^{-1} \, ,
\label{eq:probability}
\end{equation}
where $S(L)$ is the action of the path
\begin{equation}
S(L)= \frac{1}{\hbar} \int_{s_{\rm in}}^{s_{\rm out}} ds \sqrt{2 \mathcal{M}_{\rm eff}(s) 
\left[ V_{\rm eff}(s) - E_{\rm in} \right]}\, .
\label{eq:action}
\end{equation}
In Eq.~\eqref{eq:action}, $M_{\rm eff}(s)$ and $V_{\rm eff}(s)$ are the effective inertia and
effective potential energy along the path, respectively, while  $s_{\rm in}$ and $s_{\rm out}$
characterize the starting and ending points of the trajectory. We take the former as either the (calculated)
GS or isomer with corresponding energy $E_{\rm in}$,
while $s_{\rm out}$ is either (a) a semiclassical outer turning point at the same energy or (b)
a configuration consisting of two distinct fragments\footnote{In practice, we use the
expectation value of the neck operator to distinguish between compact and scissioned configurations~\cite{perez2017}.}.
Minimising the action then leads to the least action path (LAP) with maximal tunneling probability $P_{\rm max}$
and an estimate for the SF half-life:
\begin{equation}
 t_{1/2}^{\rm SF}=\frac{\ln{2}}{n P_{\rm max}},
 \label{eq:halflife}
\end{equation}
where the value of $n = 10^{20.38}$ s$^{-1}$, is based on an estimate of the relevant
timescale~\cite{Baran1978}. Another useful definition is the minimum energy
path (MEP) corresponding to the most likely path under the assumption that the effective inertia
is constant, i.e. $\mathcal{M}_{\rm eff}(s) \approx \mathcal{M}_{\rm eff}$.

As in many previous studies~\cite{Berger1989,cwiok1996,bender1998,burvenich2004,Staszczak09,Abusara10,Rodriguez14,Schunck14,Baran15,Giuliani18,Bernard2019,ryssens2019,Scamps19,agbemava2017,taninah2020}, we use EDF calculations to provide $V_{\rm eff}$, $\mathcal{M}_{\rm eff}$ and $E_{\rm in}$ as a function
of a few collective variables. An appropriate choice for the latter is essential; we take the three mass multipole moments that characterize the nuclear density of lowest order: $\beta_{20}, \beta_{22}$ and $\beta_{30}$\footnote{We define $\beta_{\ell m} \equiv 4\pi Q_{\ell m}/3R_0^\ell A$ where $Q_{\ell m}$ is a spherical harmonic and $R_0 = 1.2A^{1/3}$ fm~\cite{ryssens2019}. Note that the $\beta_{\ell m}$ are sensitive to the entire nuclear volume.}. The first quadrupole moment $\beta_{20}$ and its octupole deformation $\beta_{30}$ - i.e. the nucleus left-right asymmetry - are included in nearly all studies so far since the former is a practical way to drive the nucleus to scission while the effect of the latter on the barriers of nuclei close to stability can be 10 MeV or more~\cite{Johansson61}. The second quadrupole moment $\beta_{22}$ characterizes the triaxial deformation of the nuclear density; its inclusion forces us into a three-dimensional
representation of the nucleus. Independently of the type of EDF employed, allowing for triaxial deformation lowers the inner barrier and outer barrier of actinide nuclei by up to several MeV and  more than 0.5 MeV, respectively, yet leaves isomer excitation energies unaffected~\cite{Staszczak09,Abusara10,Ryssens23}.

We map the collective space or, more colloquially, the potential energy surface (PES) through
    Hartree-Fock-Bogoliubov (HFB) calculations constrained to specific values of $(\beta_{20}, \beta_{22}, \beta_{30})$ with MOCCa~\cite{RyssensThesis}, a tool capable of accurately representating elongated shapes that combine triaxial and octupole deformation~\cite{Ryssens2015}. The calculated total energy $E_{\rm tot}$
    already contains a rotational and vibrational correction that has consistently been included in
    the parameter adjustment~\cite{ryssens2022}; we set $V_{\rm eff} = E_{\rm tot}$ and do not subtract any additional zero-point contribution to the energy. We compute the effective inertia $\mathcal{M}_{\rm eff}$
    microscopically from the adiabatic time-dependent HFB (ATDHFB) collective inertia tensor evaluated with the cranking approximation~\cite{Baran2011,giuliani2018a}. For odd nuclei\footnote{To simplify terminology, we will refer to even-even nuclei as even nuclei, and to odd-mass and odd-odd nuclei collectively as odd nuclei throughout this work.}, we use the equal-filling approximation and search at each point on the PES for the
    blocked configuration with lowest energy after convergence, irrespective of its quantum numbers~\cite{Ryssens23}.

\begin{figure*}[htbp]
\begin{center}
\includegraphics[width=0.92\textwidth]{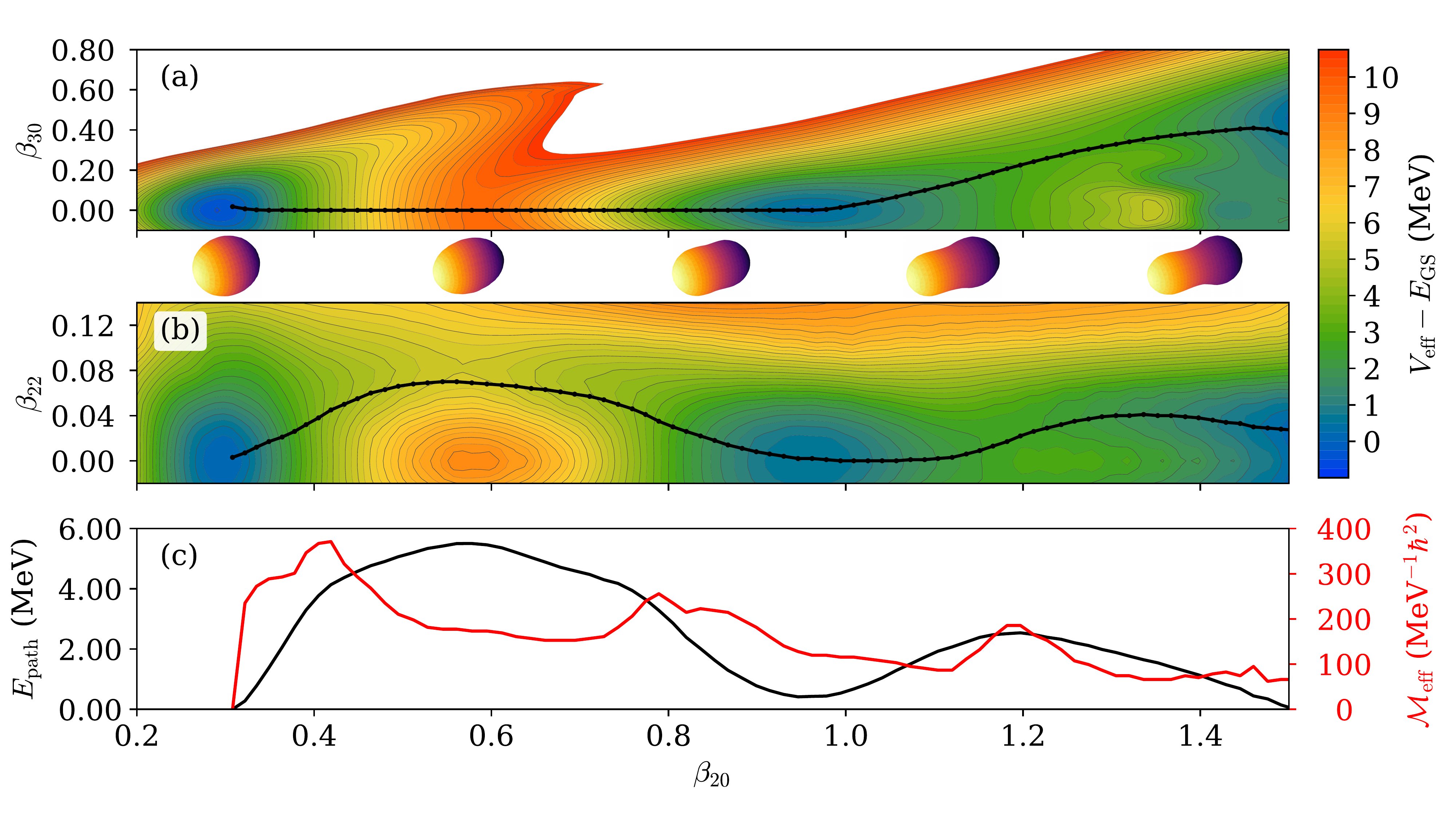}
\end{center}
\caption{\label{Fm254_PES}
PESes of $^{254}$Fm in the (a) $(\beta_{20}, \beta_{30})$ and (b) $(\beta_{20}, \beta_{22})$ planes, including the MEP and LAP, respectively. Both paths extend from the GS to the exit point $V_{\rm eff}=E_{\rm GS}$. Panel (c) contains the energy along the LAP, $E_{\rm path}(\beta_{20})=V_{\rm eff}(\beta_{20})-E_{\rm GS}$, and the corresponding effective inertia. The figure also illustrates 
3D iso-density surfaces of the nuclear configuration at various
 points along the LAP. The colours of these shapes are purely illustrative.}
\end{figure*}

The EDF-based construction of collective spaces with three or more dimensions \cite{sadhukhan2013,sadhukhan2016,flynn2022,matheson2019} 
is possible but prohibitively expensive for our purposes; instead, we
employ a two-step scheme. First we construct a 2D PES in $\beta_{20}$ and $\beta_{30}$, restricting the nucleus
to be axially symmetric; by constructing the MEP in this subspace, we find the energy-optimal octupole deformation as a function of elongation, $\beta^{\rm opt}_{30}(\beta_{20})$. In a second step, we find the LAP - hence, the SF half-life - in the two-dimensional $(\beta_{20}, \beta_{22})$ space but constrain the nucleus octupole deformation to $\beta^{\rm opt}_{30}(\beta_{20})$.
This scheme is inspired by the following observations of the archetypical double-humped barrier of an actinide nucleus : First, octupole deformation has a large effect on the energy 
near the outer barrier and the octupole matrix elements of the inertia tensor tend to be smaller than quadrupole ones, such that the LAP will remain close to $\beta_{30}^{\rm opt}(\beta_{20})$. Secondly, in the 3D space, the energy-optimal octupole deformation is reasonably independent of $\beta_{22}$
at fixed elongation $\beta_{20}$. Although this scheme is not entirely general, 
it allows us to move beyond a two-dimensional treatment where past large-scale 
studies were limited to a single collective variable~\cite{Goriely07,Giuliani18b}.

In practice, we obtain the paths - MEP in step 1 and LAP in step 2---with the nudged elastic band method implemented in the PyNEB package~\cite{flynn2022}. We illustrate this scheme with Fig.~\ref{Fm254_PES} for $^{254}$Fm: panel (a) shows the MEP in ($\beta_{20},\beta_{30}$) plane while (b) shows the LAP in the ($\beta_{20},\beta_{22}$) plane constructed in step 2. In Fig.~\ref{Fm254_PES}(c) we show the two key ingredients entering the calculation of the SF half-lives: the effective inertia and effective potential normalized w.r.t. the GS energy. From Fig.~\ref{Fm254_PES}(b), we observe that the LAP does not necessarily follow the path with the lowest potential barrier, but rather the path corresponding to the fastest decay. Rapid variations in $\beta_{22}$ induce peaks in $\mathcal{M}_{\rm eff}$, as shown in Fig.~\ref{Fm254_PES}(c), thus disfavoring triaxial deformations near the fission barriers due to their high inertia cost.

\begin{figure}[htbp]
\begin{center}
\includegraphics[width=0.9\columnwidth]{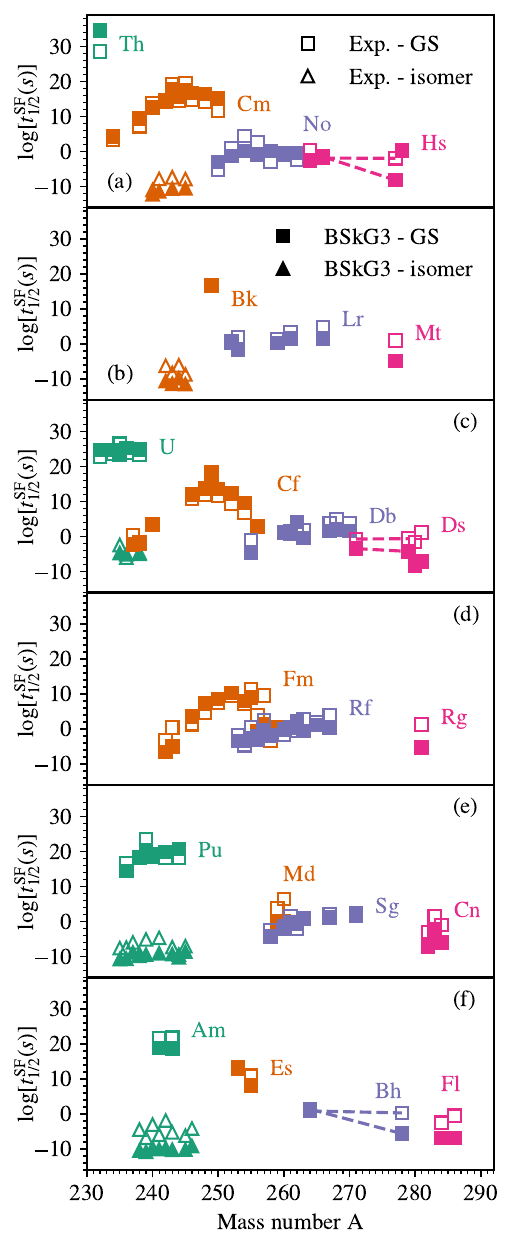}
\end{center}
\caption{\label{spf_gs_isom}
Comparison of calculated SF half-lives (filled markers) and experimental data 
(empty markers) \cite{Kondev21} for both GS (squares) and isomeric 
decay (triangles). The separation into panels and color-coding of isotopic
chains only serve the visualisation of results. We
provide all SF lifetimes in the supplementary material.}
\end{figure}

NUDAT~\cite{Kondev21} lists 107 GS and 29 isomeric SF 
half-lives for nuclei with $232 \leq A \leq 286$ nucleons. 
Fig.~\ref{spf_gs_isom} compares our results to the entirety of this compilation 
on a decimal logarithmic scale. Globally speaking, the agreement with experiment
is more than satisfactory: the model is able to match qualitatively the entire 
range of experimental data spanning nearly 60 orders of magnitude, even if 
large deviations for individual decays appear. The GS SF decay of
$^{284-286}$Fl, $^{280-281}$Ds and $^{232}$Th are the worst offenders; we will
present a more detailed analysis elsewhere~\cite{Sanchez25}, but we mention that
our $^{232}$Th PESes---like other results~\cite{Schunck14,Bernard2020}---feature a large jump in the total energy near scission.


To enable a more quantitative discussion, we define the  
rms and mean deviations with respect to $N_{\rm d}$ data points with a half-life $t^{\rm exp}_{1/2}$ as

\begin{subequations}
\begin{align}
f_{\rm mean} &= \exp \left[ \frac{1}{N_{\rm d}} \sum_{i=1}^{N_{\rm d}} \ln \frac{t^{\rm th}_{1/2,i}}{t^{\rm exp}_{1/2,i}}  \right],
\label{eq_erms} \\
f_{\rm rms} &= \exp  \left[ \frac{1}{N_{\rm d}} \sum_{i=1}^{N_{\rm d}} \left( \ln \frac{t^{\rm th}_{1/2,i}}{t^{\rm exp}_{1/2,i}} \right) ^2 \right] ^{1/2}.
\label{eq_frms}
\end{align}
\end{subequations}

Table~\ref{tab:deviations_gs_isom} lists these values for BSkG3 separately 
for GS and isomeric decays across all nuclei as well as values restricted to 
even or odd systems, respectively. BSkG3 describes 
GS and isomeric decays within less than four order of magnitude - $f_{\rm rms} \leq 5.3 \times 10^3$ - 
even if the model on average underestimates SF half-lives by about two (GS) 
to three (isomer) orders of magnitude. For even nuclei only, a significantly better agreement is found when compared to odd systems.
More interestingly, when distinguishing between long-lived ($t_{1/2}^{\rm SF} > 1$ s) and short-lived ($t_{1/2}^{\rm SF} \leq 1$ s) nuclei, we did not find a significant difference in the deviations. Both sets of nuclei exhibit an overall underestimation of 3 orders of magnitude, although $f_{\rm rms}$ is slightly larger for the short-lived cases: $1.6 \cdot 10^3$ compared to $9.6 \cdot 10^2$. The distinction is motivated by the r-process itself: only short-lived isotopes significantly influence fission recycling and final abundance patterns, whereas long-lived isotopes have negligible impact~\cite{goriely2015}.

\begin{table}[t]
    \centering
    \caption{\label{tab:deviations_gs_isom} Mean ($f_{\rm mean}$) and rms ($f_{\rm rms}$) 
    deviations (see text) for GS and isomer half-lives as obtained with BSkG3.}
    \resizebox{\columnwidth}{!}{%
    \begin{tabular}{lccccccccc}
        \hline\hline
         & \multicolumn{3}{c}{Ground State} &&&& \multicolumn{3}{c}{Isomer} \\
        & $N_d$ & $f_{\rm mean}$ & $f_{\rm rms}$ &&&& $N_d$ & $f_{\rm mean}$ & $f_{\rm rms}$ \\
        \hline
        All nuclei     & 107  & 7.5$\cdot10^{-2}$ & 1.1$\cdot10^{3}$ &&&&  29 & 7.2$\cdot10^{-4}$ & 5.3$\cdot10^{3}$ \\
        Even      & 57   & 1.4$\cdot10^{0}$  & 4.1$\cdot10^{2}$ &&&&   6 & 1.7$\cdot10^{-1}$ & 3.3$\cdot10^{1}$ \\
        Odd & 50   & 2.7$\cdot10^{-3}$ & 2.9$\cdot10^{3}$ &&&&  23 & 1.7$\cdot10^{-4}$ & 1.3$\cdot10^{4}$ \\
        \hline\hline
    \end{tabular}%
    }
\end{table}

This set of calculated SF half-lives is unique both in extent and accuracy;
although the literature on the calculation of SF half-lives with EDFs is 
extensive~\cite{erler2012,sadhukhan2013,sadhukhan2014,Baran15,zhao2016,rodriguez2017a,
lemaitre2018,rodriguez2018,matheson2019}, the only comparable set of
results that we are aware of is Ref.~\cite{Giuliani18b}. Its authors reach $f^{\rm BCPM}_{\rm rms} = 1.9\cdot10^{4}$ 
and a mean deviation of $f^{\rm BCPM}_{\rm mean} = 1.2\cdot10^{-1}$ for the 
107 known GS decays based on calculations in a 1D collective subspace using the BCPM EDF. However, this performance was 
reached only through an \emph{a posteriori} renormalisation of the collective 
inertias; the unrenormalised results can deviate from experiment by more than $20$ orders
of magnitude. BCPM performance for SF decay is also not matched with accuracy for
GS properties; it features an rms deviation of 1.6 MeV w.r.t. over 500 experimental masses of even nuclei~\cite{baldo2013}.

Also non-EDF-based studies tackling all 136 known
   SF half-lives are rare. Among mic-mac approaches, several works~\cite{baran1981,moller1994,moller2009}
   mention deviations of ``about two/three orders of magnitude'' for restricted sets
   of nuclei. The largest compilation in this category
   that we are aware of is Ref.~\cite{Blanco23}: the authors describe the
   SF GS decay of 50 even nuclei with an accuracy that is comparable to ours,
   $f^{\rm LS}_{\rm rms} = 2.8 \cdot 10^{3}, f^{\rm LS}_{\rm mean} = 0.71$,
   using the Lublin-Strasbourg liquid drop model\cite{Pomorski03} with some
   parameters tuned to this region. In comparison, BSkG3 half-lives for the 57 even nuclei in our sample (see Table~\ref{tab:deviations_gs_isom}) gives a better agreement with data with an $f_{\rm rms} = 4.1 \cdot 10^{2}$ deviate 7 times smaller. To the best of our knowledge, only local highly-parametrized formulas dedicated to SF half-lives can reach accuracies
   of about one order of magnitude~\cite{yuan2024}.

Many more studies - EDF-based or not - discuss barrier heights at length; we will present
   a detailed comparison of our predictions for this quantity with the literature elsewhere~\cite{Sanchez25}.


To gauge the robustness of our results, we calculated the SF half-lives for BSkG2~\cite{Ryssens23} and BSkG4 \cite{Grams25};
   BSkG1 \cite{Scamps2021} is excluded since its parameter adjustment did not include data relevant to fission. The
   corresponding $f_{\rm rms}$ and $f_{\rm mean}$ are listed in Table~\ref{tab:bskg_tab_comp}. Both BSkG2 and BSkG4 perform
   far worse than BSkG3 with $f_{\rm rms}$ being two and one order of magnitude larger, respectively. It is however
   not easy to link this performance difference to specific model deficiencies since the description of fission
   is sensitive to essentially all aspects of an EDF parameterization~\cite{samyn2004,jodon2016,bulgac2018,ryssens2019,guan2021,dacosta2023}. The reduced accuracy of BSkG2 is likely, at
   least partially, due to its objective function that included only 12 RIPL-3 barriers, as opposed to BSkG3/4. Compared
   to BSkG3, the most recent model predicts weaker pairing correlations; this leads to larger inertias and a corresponding increase in
   the half-lives that translates to BSkG4 overestimation of the data on average. 
   
   In view of the sensitivity of the SF half-lives, the differences between models should be considered as small even if the
   $f_{\rm rms}$ vary by two orders of magnitude; BSkG2 and BSkG4 still compare favourably to the
   literature on SF half-lives. Although a complete uncertainty quantification for EDF-based fission modelling is out of reach today~\cite{agbemava2017}, our results clearly establish that our fitting protocol produces models that
   combine excellent accuracy for GS and fission properties in a robust way.


\begin{table}[h]
    \centering
    \caption{\label{tab:bskg_tab_comp}Mean ($f_{\rm mean}$) and rms ($f_{\rm rms}$) 
    deviations of the 107 (57 even and 50 odd) GS SF half-lives obtained with BSkG2/3/4.}
    \resizebox{\columnwidth}{!}{%
    \begin{tabular}{l|ccc|ccc}
        \hline
        \hline
        & \multicolumn{3}{c}{$f_{\rm mean}$} & \multicolumn{3}{c}{$f_{\rm rms}$} \\
        & BSkG2 & BSkG3 & BSkG4 & BSkG2 & BSkG3 & BSkG4 \\
        \hline
        All            & 6.5$\cdot10^{-4}$ & 7.5$\cdot10^{-2}$ & 1.4$\cdot10^{2}$  & 1.5$\cdot10^{5}$ & 1.1$\cdot10^{3}$ & 3.7$\cdot10^{4}$ \\
        Even      & 2.3$\cdot10^{-2}$ & 1.4$\cdot10^{0}$  & 2.3$\cdot10^{3}$  & 2.5$\cdot10^{4}$ & 4.1$\cdot10^{2}$ & 4.1$\cdot10^{4}$ \\
        Odd & 1.1$\cdot10^{-5}$ & 2.7$\cdot10^{-3}$ & 5.6$\cdot10^{0}$  & 9.0$\cdot10^{5}$ & 2.9$\cdot10^{3}$ & 3.3$\cdot10^{4}$ \\
        \hline
        \hline
    \end{tabular}
    }
\end{table}

We presented here our modelling of SF half-lives: after obtaining two two-dimensional potential energy surfaces and the associated collective inertia with the BSkG3 parameterization for each nucleus, we calculate its fission half-life from the LAP that connects the nuclear GS to a fissioned configuration. Despite the computational complexity, we applied this strategy to all known SF half-lives and found that our results reproduce the 107 GS and 29 isomeric decay rates within four orders of magnitude.

To the best of our knowledge, this is the first time an entirely microscopic description of fission has reached this accuracy without introducing additional renormalisation parameters; BSkG3 is the first EDF-based model that can match microscopic-macroscopic models in this arena. We are hopeful that the quality extends to the calculation of - for instance - neutron-induced fission rates since BSkG3 reproduces the 45 available empirical fission barrier heights of RIPL-3 with unprecedented accuracy, even if the latter are model-dependent observables~\cite{Grams23}.  Strikingly,  BSkG3 demonstrates that this performance for half-lives can be combined with state-of-the-art accuracy for GS observables such as masses and charge radii, all while maintaining a realistic equation of state that is compatible with the existence of heavy ($M > 2 M_{\odot}$) neutron stars. By combining all of these feats, BSkG3 sets a new standard in data generation for r-process simulations.

Nevertheless, our description of SF is not without issues. The first is the remaining systematic deviation between our calculations and experimental data. The second concerns odd systems: whether adiabatic evolution in a small collective space is a reasonable description for such systems is an open question~\cite{Bender20}. Third is the generality of the collective space: both hexadecapole and pairing degrees of freedom are relevant~\cite{Bernard2019,Guzman2022,Lay2024} even if we did not include them to keep the calculations feasible. Fourth, our perturbative treatment of the inertia tensor can be made more microscopic \cite{Baran2011,Giuliani18}.

We will focus on the extension of these results to (i) thousands of neutron-rich nuclei and (ii) fission reaction rates with TALYS~\cite{Koning23}. Further work on fragment yields should render the BSkG3 set of fission data complete for r-process simulations. An investment in machine-learning techniques should bring down the computational cost for building PESes dramatically, allowing us to tackle the large-scale description of fission properties more systematically and perform statistical uncertainty quantification~\cite{Lay2024}. Finally, our work opens possibilities for the refinement of future models: a baseline accuracy of a few orders of magnitude for SF half-lives should be sufficient to allow their systematic inclusion in parameter adjustments. Replacing in this way the RIPL-3 barriers in the BSkG parameter adjustment is desirable because barriers are model-dependent observables and the experimental data on SF half-lives spans a larger range of $N$ and $Z$.

%
\textit{Acknowledgments}---We are grateful to G. Grams for his help during the start-up phase of this project, to E. Flynn, D. Lay and K. Godbey for their guidance in the use of PyNEB and to S. Giuliani and L. Robledo for extensive discussions on collective inertias and for providing us with the calculated BCPM barriers and half-lives. S.G. and W.R. acknowledge financial support from F.R.S.-FNRS (Belgium). This work was supported by the Fonds de la Recherche Scientifique - FNRS and the Fonds Wetenschappelijk Onderzoek - Vlaanderen (FWO) under the EOS Project No O000422 and O022818F. S.B. acknowledges the support of the FWO fellowship for fundamental research (contract no. 1167324N).
This research benefited from computational resources made available on the Tier-1 supercomputer Lucia of the Fédération Wallonie-Bruxelles, infrastructure funded by the Walloon Region under the grant agreement nr 1117545.
Further computational resources have been provided by the clusters Consortium des Équipements de Calcul Intensif (CÉCI), funded by F.R.S.-FNRS under Grant No. 2.5020.11 and by the Walloon Region.
\appendix
\section{Content of the supplementary material}

The supplementary material includes two data files containing both the experimental and BSkG3 spontaneous fission (SF) half-lives for ground states (\texttt{sf\_hf.gs.BSkG3.dat}) and isomers (\texttt{sf\_hf.is.BSkG3.dat}) discussed in this work. Each file contains not only those with definite experimental values but also those for which only experimental limits (upper or lower bounds) are available. The data columns are organized as follows: the first and second columns list the proton number $Z$ and mass number $A$, respectively. Columns three and four provide the experimental SF half-lives and their associated uncertainties. Column five indicates the type of experimental data---whether it is a measured value (\texttt{equ}), an upper limit (\texttt{max}), or a lower limit (\texttt{min}). Only the \texttt{equ} values were used in the main analysis. Column six gives the reference source of the experimental data, and the final column contains the corresponding BSkG3 predictions.

Although most of the data were obtained from Ref.~\cite{Kondev21}, SF half-lives for 10 isomeric states were taken from the IAEA's LiveChart of Nuclides online tool~\cite{LiveChart}.

\bibliographystyle{elsarticle-num} 
\bibliography{apssamp}
\end{document}